# Comment on the Evidence of isostructural phase transitions in elemental zirconium

Joseph Gal*

Ilse Katz Institute for Nanoscale Science and Technology ,
Ben-Gurion University of the Negev, Beer Sheva ,84105 Israel

*jgal@bgu.ac.il

## Abstract

It is argued that the article by O'Bannon et al. "**High pressure stability of β-Zr: no evidence for isostructural phase transitions**" published recently (July 2021) in High Pressure Research has no experimental foundation and the statement "no evidence" is misleading. The existence of the transitions ω → β → β' have been experimentally observed by Akahama et al. (2001) DAC experiment using gold PTM. The experiment by E. Stavrou et al. , refined by J. Gal, confirmed cascading transitions ω → β → β' in compressed Zr metal powder with no PTM (Zr PTM), confirming the isostructural bcc cascade β → β'→ β".

It is demonstrated that the observation of the ω → β → β' depend on the selected pressure transmitting media used in the DAC.

Contrary to O'Bannon et al. article it is concluded that Zr metal, at the pressure range of ~ 30 - 200 GPa, exhibits a meta- stable ω phase, which transforms by a first order phase transition to β phase, which then transforms to the β" ground state via the β' phase. Isostructural phase transition has been reported only in Ce metal, thus, the β-Zr → β-Zr isostructural phase transitions has fundamental interest in condensed matter physics.

**Introduction**- Just recently E. F. O'Bannon et al. published in High Pressure Research (July 2021) [1] the article "High pressure stability of β-Zr: no evidence for isostructural phase transitions". Their statement "no evidence" is based on Zr DAC experiments performed with Ne pressure transmitting medium (PTM). In 1991, however, the isostructural ω →bcc→bcc' phase transition at ~58 GPa in elemental Zr was observed by Akahama et al. [2] using a diamond anvil cell (DAC) with a gold PTM. Later on (2018), Stavrou et al. [3] performed a precise XRD structural study of Zr metal powder without a PTM. This experiment, up to 210 GPa at ambient temperature, reconfirmed the existence of the isostructural bcc-to-bcc' first-order phase transition claimed by Akahama et al.. In a recent contribution [4] accurate fittings of the data reported by Stavrou et al. using Vinet (VIN) or Birch-Murnagham (BM) equations of state (EOS), revealed the existence of another bcc phase with different elastic properties in addition to the reported β and β' phases. It was shown that the first order volume collapse at ~58GPa (β → β') is followed by a moderate transition to a bcc-β" phase where the β' phase is stable up to 110 GPa. Above 110 GPa the bcc-β" is dominant and stable up to at least ~220 GPa. Although both Akahama et al. and Stavrou et al. used different PTMs, the existence of the isostructural bcc-to-bcc phase transition was clearly observed.

Further on, Pigott et al. (2020) confirmed the existence of the bcc Zr-β phase. By using hydrostatic helium PTM [5], the ω → β transformation erupted at ~68 GPa. The experiment reported by E.F. O'Bannon et al. [1] does not observe the ω → β transition, then obviously, the no evidence isostructural phase transitions in compressed Zr could be claimed. Therefore, the non-evidence statement has no experimental foundations. The observation of the ω → β → β' isostructural phase cascade is dependent on the chosen PTM. Under such pressure conditions the β → β' transition cannot be denied and the statement "no evidence for isostructural phase transitions in Zr metal" is totally misleading.

# Detailed Proofs:

## Akahama et al. (1991) EOS:

The first observation of the bcc β → β' transition reported by Akahama et al. [2] is depicted in Fig. 1(a). The DAC experiment was performed with Zr metal foils sandwiched between Au foils and compressed up to ~68 GPa. The Au foils protect the compressing diamonds and serves as a quasi-PTM, namely the Zr metal sample is compressed as if no PTM exists in the DAC, similar to the Stavrou et al. [3] experiment.

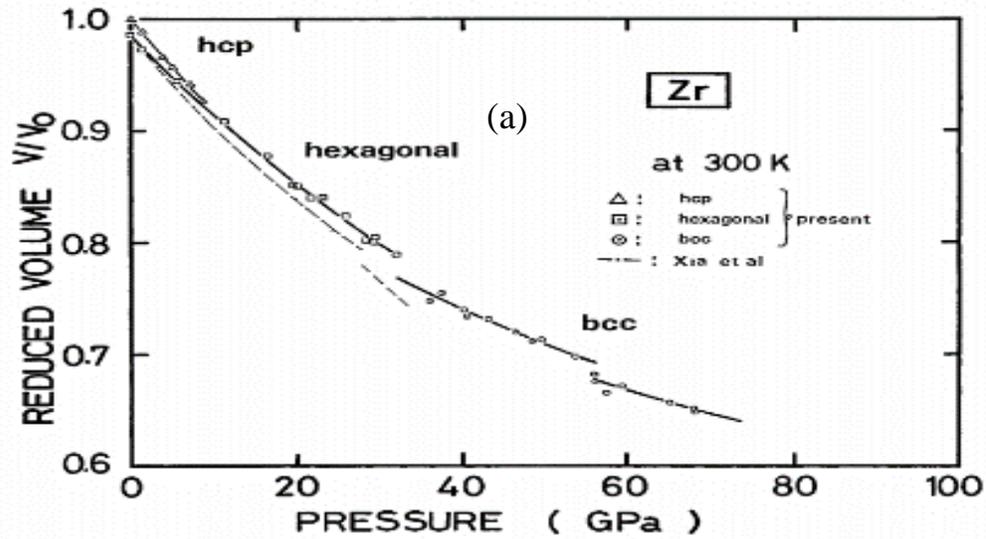

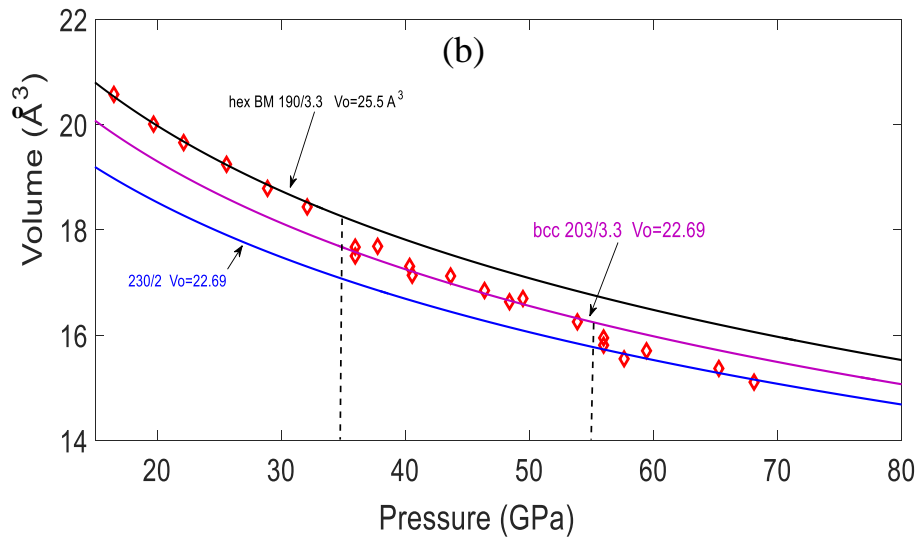

**Fig.1**: (a) Akahama et al. [2, therein Fig.3] EOS (2001). (b) The enlarged and reanalyzed BM fitting parameters are marked $B_o/B_o'$ in the figure, where $V_o$ is a fitting parameter.

## Stavrou et al. (2018) EOS:

E. Stavrou et al. [3] performed up to 210 GPa a precise XRD structural study of Zr metal powder, at ambient temperature, without a PTM. Their results reconfirmed the existence of the isostructural bcc-to-bcc' phase transition and independently suggested a first-order phase transformation. In a recent contribution by J. Gal [4], an accurate fittings of the data reported by Stavrou et al. was reported. Fitting with VIN or BM equations of state (EOS) revealed that at the pressure range ~ 58-110 GPa another bcc phase with different elastic properties exists, as shown in Fig. 2.

In conclusion, from the crystallographic point of view, there is no difference between the results reported by Akahama et al. [1] and Stavrou et al. [3], Akahama et al. measured the pressure up to ~ 70 GPa., however, E. Stavrou et al. continued the measurement up to 220 GPa, revealing the bcc-β'and the bcc-β" phases [4]. In addition, the existence of the bcc-β" phase is confirmed by the fitting of Zr melting data. By assuming a Debye solid and demanding that the derived EOS fitting parameters will simultaneously fit the melting points, good fits were obtained upon constructing the melting curve [4].

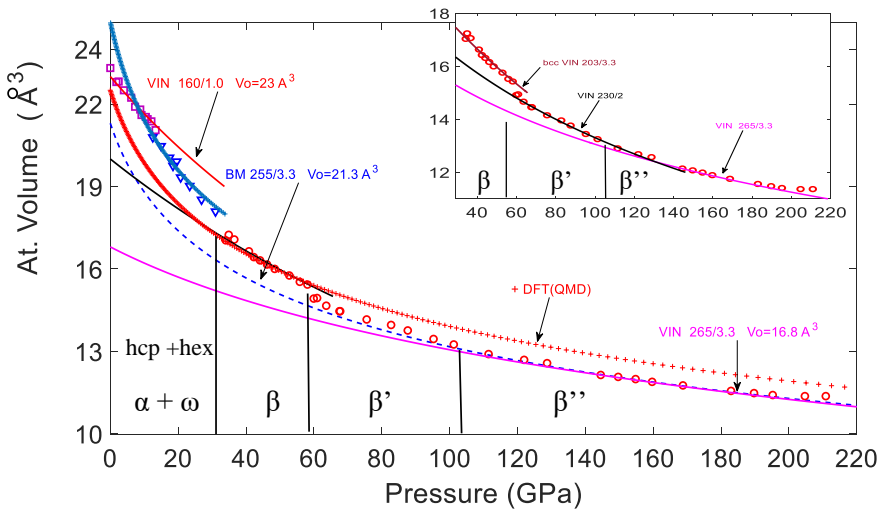

**Fig. 2**: EOS of the elemental zirconium - room temperature isotherm. The experimental data points are taken from Stavrou et al. [3] and are fitted with the VIN or BM EOS. The successive transformation from hcp to bcc structures are observed. The derived bulk moduli are assigned $B_o, B_o'$ and $V_o$ pointed with arrows. Above 30 GPa the β phase is dominant and a successive cascade to bcc-β' and bcc-β'' are clearly observed. The bcc-β structure is stable up to ~58GPa, where a first order phase transition occurs with a volume reduction of ~4% which is stable up to ~105GPa (inset black solid line). Above 105 GPa and up to 220 GPa a stable bcc-β" exists (magenta solid lines). The dashed blue line represents the BM simultaneous best fit of the experimental data in the P-V and P-T planes [4], indicating that both BM or VIN predict the β → β' isostructural phase transition. The red + sign is the DFT(QMD) simulation which match only the experimental β phase data.

pg. 4

## Pigott et al. (2020) EOS:

J.S. Pigott et al. [5], re-measured the Zr 300 K isotherm under hydrostatic (He-PTM) conditions and re-confirmed the existence of the ω → β first order collapse, evidencing the existence bcc β-zirconium phase at ~ 68 GPa. These results are depicted in Fig.3, marked by red sticks.

Eventually, the citation of Pigott et al. by O'bannon et al. [1] (Fig.1, therein) as "no evidence" did not take in to account the hydrostatic He- PTM compression, which reveals the ω → β transition. Up to 80 GPa, O'Bannon et al. [1] measurements did not observe this transition when using Ne PTM.

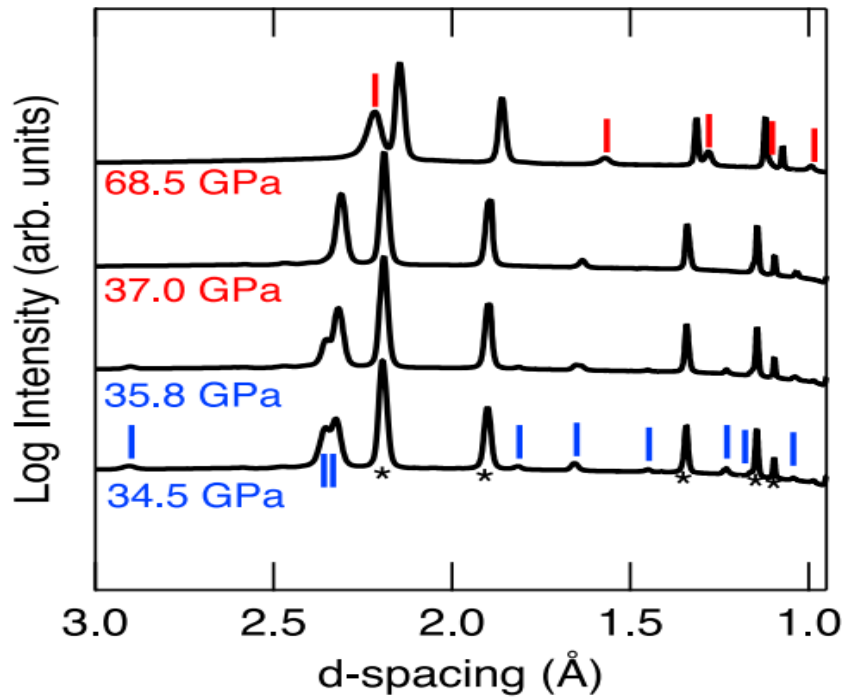

Fig. 3: XRD measurements at ambient temperature of elemental Zr showing ω → β transition under hydrostatic (He PTM) compression. The red sticks denote the bcc phase. The ω → β transition is clearly observed around 68.5 GPa.

## Simone Anzellini et al. (2020) :

O'Bannon et al. [1] argued that within the statistical error the claimed β → β phase does not exist quoting and acknowledging Anzellini et al. [6]. Quotation from Anzellini shows completely different situation:

"We suggest that non-hydrostatic compression effects (pressure gradients and/or lattice strains) explains the observation of an isostructural transition to **β**-Zr with a 4% volume collapse at 58 GPa . The apparent isostructural **β**-Zr → **β**-Zr transition could be caused by a change in stress distribution in the course of the compression, for instance, due to deformation ("cupping") of diamond anvils".

The suggestion posed in last sentence can be applied to any the DAC measurement containing PTM, raising the question whether all the interpretations of DAC measurements, EOSs and melting points, are exposed to cupping.

## O'Bannon et al. (2021):

O'Bannon et al. [1] argue that within the statistical error the claimed β → β phase does not exist. The analysis of any experimental point refers to the center of mass of the error bar. What should be looked at is the trend of the experimental points relative to the fitting lines (Fig. 2). In the present case all points in the region from 59 to 110 GPa are above the BM or VIN fitting lines. Namely, both reveal the β' phase shown by Akahama and Stavrou. In addition, O'Bannon et al. performed three runs on Zr foil with a Ne PTM, labelled 1, 2, 3. As shown in Fig. 4 (green and red points) up to 80 GPa no evidence of phase transition was observed. Wrongly quoting Pigott et al. (2020), as at 68.5 GPa the β phase abruptly erupts (see above). By quoting Anzellini et al. [6] as evidence for their claim, O'Bannon et al. further mismatch Anzellini et al. statement that "the apparent β-Zr → β-Zr transition could be caused by a change in stress distribution in the course of compression, for instance due to the deformation of the DAC".

It should be noted that the DAC measurements with Ne PTM presents a completely different type of experiment and should not be compared to experiments with different PTMs. The PTM plays a crucial role in all the DAC experimental results.

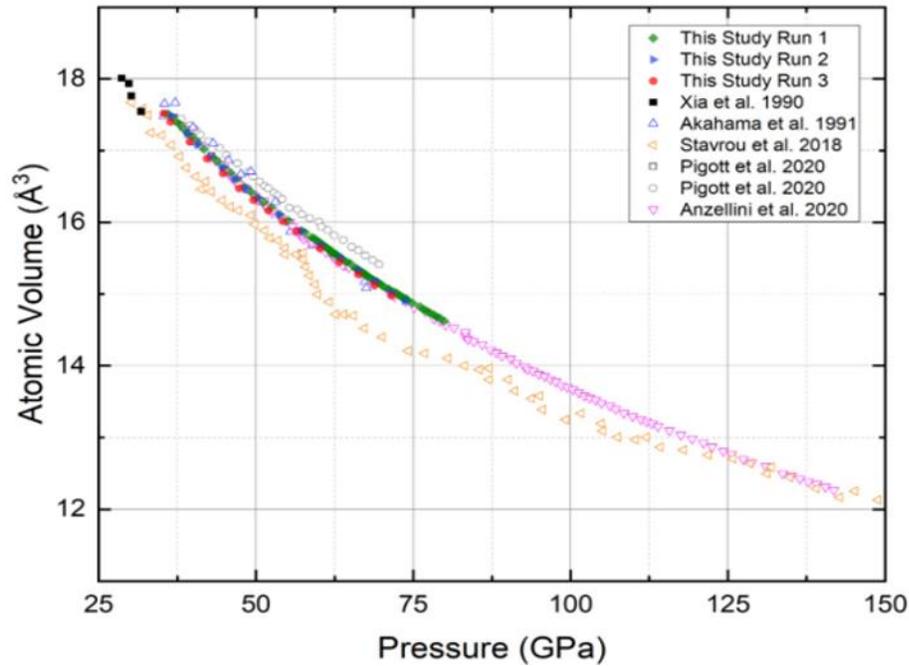

**Fig. 4**: Cut and paste from Anzellini et al. [6] (therein, Fig. 5). The three runs 1, 2, 3 with Ne PTM indicate pressure extension up to ~ 82 GPa (green diamonds). The ω → β phase transition is not observed. Citing Pigott et al. (2020) is wrong, as by using He PTM at 68.5 GPa, the β-phase erupts.

## Discussion

No doubt that the isostructural phase transition has fundamental interest in condensed matter physics. Up to now, isostructural phase transition had been observed only in Ce metal [7]. The observation β → β' transition by Akahama et al. [2], confirmed by Stavrou et al. [3], should be related to the pressure conditions developed in the DAC with Zr or Au PTM. Anzellini et al. [6] suggested that the isostructural β-Zr → **β-Zr** transition could be caused by a change in stress distribution in the course of the compression, for instance, due to deformation of the diamond anvil which even includes the gasket. Such a deformation can be claimed for any DAC experiment using less compressible PTMs like for example MgO or KCl. Nevertheless, the existence of the β-Zr → **β-Zr** transition is an experimental fact observed by two independent experiments [2,3].

In order to prove their hypothesis "**no evidence for β-Zr → β-Zr isostructural phase transitions**", O'Bannon et al. presented results from DAC experiments extended to ~ 80 GPa performed with a Ne PTM (see Fig.4). Their experimental results do not differ from the results reported by Anzellini et al., results using Ne PTM. The fact that the ω → β phase transition is not observed when using Ne PTM indicate an extremely compressible PTM that should be related to the Ne low bulk modulus $B_o$= 13(2) GPa [8]. So soft, that solid Ne PTM behaves close to a liquid [9]. In other words, the softness of the Ne PTM seems to be the reason of the non-observance of the ω → β first order transition, similar to cases of the transition metals [13]. It should be noted that the transition ω → β pressure transition points are strongly dependent on the chosen PTM: KCl- 3 GPa [10], Zr -30(2) GPa, Au- 35(2) GPa and He- 68(2) GPa.

O'Bannon et al. [1] experiment never reached the β phase as shown in Fig. 4, raising the question on what experimental evidence they deduced the superlative "no evidence". There is no logic in comparing the experiments performed with Ne PTM to experiments using Zr or Au PTMs. The experiment of Pigott et al. [5], performed with hydrostatic He- PTM (Fig.3), shows clearly the first order ω → β volume collapse which is not observed when using Ne PTM. Hui Xia et al. [11] were the first to report the observation of the ω → β transition (1990). A recent proof for ω → β transition in elemental zirconium at 3 GPa, has been also reported by P. Parisiades et al. [10] using KCl PTM.

It is hard to understand how O'Bannon et al. came to the conclusion that no evidence for β-Zr → β-Zr isostructural transition exist just from an experiment where the β phase has not been reached. The present comment calls for further DAC experiments with less compressible PTMs like for example MgO or $Al_2O_3$. The existence of the β-Zr → β'-Zr invites first principles volume collapse analysis [7]. The PTM plays a crucial role in all DAC experimental results [12] which is also shown in the presents Zr case.

The statement "**no evidence for β-Zr → β-Zr isostructural phase transitions**" based on Ne PTM has no experimental foundations and is misleading.

## Conclusions

The statement "no evidence for isostructural phase transitions in Zr metal" claimed by O'Bannonet al. [1] contradicts two independent experimental evidences [2,3]. The DAC experiment with Ne PTM leads to denying the existence of the ω → β first order phase transition which was reported by five independent experiments [2,3,5,10,11].

The fact that the transition ω → β is not observed using Ne PTM in DAC experiment indicate an extremely compressible PTM ($B_o$= 13(2) GPa [8]), where the solid Ne PTM behaves very close to a liquid.

The "**no evidence for β-Zr → β-Zr isostructural phase transitions**" declaration has no experimental evidences and is misleading.

It can't be denied that Zr metal at room temperature in the pressure range of 30 - 200 GPa exhibits a meta-stable hexagonal ω phase which collapses to a bcc β phase, which then transforms the to the bcc β" ground state via the bcc β' phase.

## References

[1] E. F. O'Bannon , P. Söderlind, D. S. Sneed, M. J. Lipp, H. Cynn, J. S. Smith, C. Park and Zs. Jenei,

High Pressure Research, 30 July 2021, doi:10.1080/08957959.2021.1957863

"High pressure stability of β-Zr: no evidence for isostructural phase transitions"

[2] Y. Akahama, M. Kobayashi and H. Kawamura,

    J. of The Physical Society of Japan, Vol.60 No.10 3211(1991)

  "High-Pressure X-Ray Diffraction Study on Electronic $s$-$d$ Transition in Zirconium"

[3] Elissaios Stavrou, Lin H. Yang, Per Söderlind, Daniel Aberg,, Harry B. Radousky,1 Michael R. Armstrong,, Jonathan L. Belof, Martin Kunz, Eran Greenberg, Vitali B. Prakapenka, and David A. Young,

PRB Rapid Communications 98, 220101(R) (2018).

[4] J. Gal , Physica B: Condens Matter, 613,15 July 2021, 412979.

"Cascading crystallographic transitions α → ω → β → β'→ β" and melting curve of elemental zirconium"

[5] J. S. Pigott , Nenad Velisavljevic, Eric K Moss, Dmitry Popov, Changyong Park , James A Van Orman, Nikola Draganic, Yogesh K Vohra and Blake T Sturtevant.

J. Phys, Condens. Matter, **32** (2020) 12LT02 (6pp).

"Room-temperature compression and equation of state of body-centered cubic zirconium"

[6] Simone Anzellini , François Bottin, Johann Bouchet, and Agnès Dewaele,

PHYSICAL REVIEW B **102**, 184105 (2020).

"Phase transitions and equation of state of zirconium under high pressure"

[7] M. J. Lipp, D. Jackson, H. Cynn, C. Aracne, W. J. Evans, and A. K. McMahan
Phys. Rev. Lett. 101, 165703 – Published 15 October 2008

"Thermal Signatures of the Kondo Volume Collapse in Cerium"

[8] J. Gal and L. Friedlander , Physica B 625 (2022) 413445.

"Melting curves and bulk moduli of the solid noble gasses He, Ne, Ar, Kr and Xenon"

[9]  K. Takemura,  High Pressure  Research, VOL. 41, NO. 2, 155–174  (2021).

"Hydrostaticity in high pressure experiments: some general observations and guidelines for high pressure experimenters" . Applying  Hooks law  $\sigma_{ij} = C_{ij}\, \varepsilon_{kl}$,  (hydrostaticity)

[10] P. Parisiades, F. Cova, and G. Garbarino, PRB. 054102 (2019)100

"Melting curve of elemental zirconium, observation of ω → β transition with KCl PTM".

[11]  Hui Xia, Steven J. Duclos, Arthur L. Ruoff, and Yogesh K. Vohra
        Phys. Rev. Lett. 64, 204 – Published 8 January 1990
"New high-pressure phase transition in zirconium metal reporting the ω → β transition"

[12] A.Dewaele, M. Mezouar, Nicolas Guignot and Paul Loubeyre,

 PRL 104, 255701 (2010).